\documentclass{amsart}

\usepackage{amsmath}
\usepackage{amsfonts,amssymb,amsmath,amsthm,mathrsfs}
\usepackage{url}
\newtheorem{thrm}{Theorem}[section]

\newtheorem{lem}[thrm]{Lemma}
\newtheorem{ex}[thrm]{Example}
\newtheorem{prop}[thrm]{Proposition}

\theoremstyle{definition}

\begin{document}
\author[C.A.Mantica and L.G.Molinari]{Carlo Alberto Mantica and Luca Guido Molinari}
\address{C.~A.~Mantica (ORCID: 0000-0001-5638-8655) and L.~G.~Molinari (corresponding author, ORCID:
0000-0002-5023-787X): 
Physics Department Aldo Pontremoli,
Universit\`a degli Studi di Milano and I.N.F.N. sezione di Milano,
Via Celoria 16, 20133 Milano, Italy.}
\email{carlo.mantica@mi.infn.it, luca.molinari@mi.infn.it}

\title[Codazzi tensors and Cotton gravity]{Codazzi tensors and their space-times\\ and Cotton gravity}
\date{16 jan 2023}

\begin{abstract}
We study the geometric properties of certain Codazzi tensors for their own sake, and for their appearance in the recent theory of Cotton gravity.\\
We prove that a perfect-fluid tensor is Codazzi if and only if 
the metric is a generalized Stephani universe. 
A trace condition restricts it to a warped space-time, 
as proven by Merton and Derdzi\'nski. We also give necessary and sufficient conditions for a space-time to host a current-flow Codazzi tensor. 
In particular, we study the static and spherically symmetric cases, which include the Nariai and Bertotti-Robinson 
metrics. The latter are a special case of Yang Pure space-times, together with spatially flat FRW space-times with
constant curvature scalar. \\
We apply these results to the recent Cotton gravity by Harada. The equations have the freedom of choosing a Codazzi tensor, that 
constrains the space-time where the theory is staged. 
The tensor, chosen in forms significative for physics, implies the form 
of the Ricci tensor, and the two specify the energy-momentum tensor, which is the source in Cotton gravity for the chosen metric. For example, the Stephani, Nariai and Bertotti-Robinson space-times solve Cotton gravity with physically sensible
energy-momentum tensors.\\
Finally, we discuss Cotton gravity in De Sitter space-times. 
\end{abstract}


\maketitle

\section{Introduction}
In Ref.\cite{Harada21} Junpei Harada proposed an extension named ``Cotton gravity'' of the Einstein equations \cite{Bargueno,Harada21b}, where
the geometric term (the Einstein tensor) is replaced by the Cotton tensor,
and the source (the energy-momentum tensor) is replaced by gradients of the energy-momentum:
\begin{align}
{\sf C}_{jkl} =  \nabla_j T_{kl}-\nabla_k T_{jl}- \frac{g_{kl}\nabla_j T -g_{jl}\nabla_k T}{n-1} \label{HEq}
\end{align}
$T$ is the trace $T^k{}_k$, and the Newton constant is absorbed in $T_{jk}$. The Cotton tensor
\begin{align}
{\sf C}_{jkl} =  \nabla_j R_{kl}-\nabla_k R_{jl}- \frac{g_{kl}\nabla_j R -g_{jl}\nabla_k R}{2(n-1)} \label{Cott}
\end{align} 
is related to the Weyl tensor, ${\sf C}_{jkl} = -\frac{n-2}{n-3}\nabla_m C_{jkl}{}^m$, and contains third derivatives of the metric tensor.\\
While solving \eqref{HEq} for a vacuum ($T_{kl}=0$) static spherically symmetric solution, Harada obtained a generalization of 
the Schwarzschild solution:  $$ ds^2 = -b^2(r) dt^2 + \frac{1}{b^{2}(r)} dr^2 + r^2d\Omega^2_2 $$ 
with $b^2(r)= 1-2M/r + ar + br^2$. Next, 
in Ref.\cite{Harada22}, he applied the theory to describe the rotation curves of several galaxies, where the effect of the possible dark-matter halo is
supplanted by the modified gravitational potential.

The difference of equations \eqref{HEq} and \eqref{Cott} shows that 
\begin{align}
\mathscr C_{kl} =R_{kl}-T_{kl}-g_{kl} \frac{R-2T}{2(n-1)} \label{CRT}
\end{align}
is a Codazzi tensor: 
\begin{align}
\nabla_j \mathscr C_{ik} =  \nabla_k \mathscr C_{jk} \label{CODAZZI}
\end{align} 

Equations \eqref{CRT} and \eqref{CODAZZI} are equivalent to the Harada equation \eqref{HEq} for Cotton gravity. 
In fact, with $R_{kl}=\mathscr C_{kl}+T_{kl}+g_{kl}\frac{R-2T}{2(n-1)}$ the Cotton tensor tensor \eqref{Cott} is 
constructed, and the Codazzi condition ensures that \eqref{HEq} is obtained. \\
It turns out that the third order character of \eqref{HEq} manifests in the supplemental terms provided by the Codazzi tensor
either to the Ricci tensor or to the energy-momentum tensor, or both in \eqref{CRT}.\\
The case $\mathscr C_{kl}=0$ in \eqref{CRT} restores the Einstein equations, and eq.\eqref{HEq} is identically true.
The ``trivial'' case $\mathscr C_{jk}=B g_{jk}$, adds a cosmological constant. If $\mathscr C_{kl}\neq 0$, eq.\eqref{CRT} can still be interpreted as the Einstein equation, but with a modified energy-momentum tensor:
\begin{align}
 R_{kl} - \tfrac{1}{2} R g_{kl}  = T_{kl} + \mathscr C_{kl} - g_{kl} \mathscr C^r{}_r \label{EINSTEIN+COD}
 \end{align}

Let us mention that Codazzi tensors appear in the geometry of hypersurfaces \cite{Lovelock}. 
A Lorentzian hypersurface in a Minkowski space-time has Riemann tensor 
$R_{jklm} = \Omega_{jl}\Omega_{km} -\Omega_{jm}\Omega_{kl}$ where $\Omega_{jk}$ is a Codazzi tensor. 
The trivial case $\Omega_{jk} = \frac{R}{n(n-1)} g_{jk}$ corresponds to a constant-curvature hypersurface, and
the tensor has a single, constant eigenvalue.

A non-trivial Codazzi tensor poses important limitations on the geometry of the hosting space-time. \\
Among the possible tensors, we choose to investigate two simple and physically relevant ones, that
often appear in the expressions of the Ricci or of the energy-momentum tensors. They involve the basic kinematic
quantities
$u_i$ and $\dot u_i$.  

We begin with the ``perfect fluid'' tensor $\mathscr C_{jk}=Au_j u_k +Bg_{jk}$ with the Codazzi property. 
Andrzej Derdzi\'nski \cite{Der81}
proved that if $\mathscr C^k{}_k$ is a constant, then the space-time is warped (GRW, generalized Robertson-Walker space-time), i.e. there are coordinates such that 
\begin{align}
ds^2 = -dt^2 + a^2(t) g^\star_{\mu\nu} ({\bf x}) dx^\mu dx^\nu \label{GRW}
\end{align} 
with Riemannian metric $g^\star_{\mu\nu}$.
The hypothesis was weakened by Gabe Merton \cite{Merton13}, who showed that a necessary and sufficient 
condition for the GRW space-time is $v^j\nabla_j {\mathscr C}^k{}_k=0$ for all vectors $v^ju_j=0$ (the result was proven in Riemannian signature,
but it also holds in Lorentzian).\\
In Theorem \ref{ONE} we prove that a perfect fluid 
tensor is Codazzi if and only if the space-time is ``doubly twisted'', i.e there are coordinates such that
\begin{align}
ds^2 = - b^2(t,{\bf x}) dt^2 + a^2(t,{\bf x}) g_{\mu\nu}^\star ({\bf x}) dx^\mu dx^\nu \label{DTW}
\end{align}
with the special condition that $(\partial_t\log a)/b$ only depends on time $t$. Remarkably, this metric  with the constraint happens to be a generalization of the well known Stephani Universes.\\
We discuss special cases, including Merton's result, 
and obtain the general form of the Ricci tensor.

Next we study the ``current flow'' tensor $\mathscr C_{jk}=\lambda (u_j \dot u_k +\dot u_j u_k)$ with the Codazzi condition and closed
vector field $\dot u_j$. The field $u_j$ turns out to be vorticity-free but not shear-free. This makes the metric more general than doubly-twisted, eq.\eqref{FERR}. However, if
it is constrained to be static, a useful form of the Ricci tensor is obtained. We list some of the several examples that
can be found in the literature. 

Finally, we consider Yang Pure space-times. They are characterised by a Ricci tensor that is a Codazzi tensor. Among
examples, we show that a Friedmann-Robertson-Walker metric is Yang Pure if and only if $\nabla_j R=0$.\\
This concludes Section 2 of the paper.

In Section 3 we show that these results are interesting for the Cotton gravity by Harada. If nontrivial, the Codazzi tensor
introduces geometric or unconventional matter content in the Einstein equation, depending on the point of the view, in a way different
than other extended theories of gravity.\\
This suggests a solution to the Harada equations which goes as follows: given the form of a Codazzi tensor, this determines a class of 
space-times that host the tensor. The space-time in turn determines the Ricci tensor. 
Finally, the Codazzi and the Ricci tensor in eq.\eqref{CRT} determine the energy-momentum tensor of the 
Harada equation.\\
The two Codazzi tensors that are here studied, modify the energy-momentum in its perfect-fluid
component or in the current component.\\
We end with a discussion of De Sitter space-times, for which Ferus \cite{Ferus} identified the general form of Codazzi tensors.

We employ the Lorentzian signature $(-+...+)$, latin letters for space-time components and greek letters for space components. 
A dot on a quantity $X$ is the operator $\dot X =u^k\nabla_k X$. The symbols $\eta $, $\epsilon$ are the scalar functions 
$\eta = \dot u^k \dot u_k$ and $\epsilon = \dot u^k\nabla_k\eta$.

\section{Codazzi tensors and their space-times}
In refs.\cite{RC,DS}  we showed that a Codazzi tensor always satisfies an algebraic identity with the Riemann tensor (it is ``Riemann compatible''):
\begin{align}
\mathscr C_{im} R_{jkl}{}^m +\mathscr C_{jm} R_{kil}{}^m +\mathscr C_{km} R_{ijl}{}^m =0. \label{RiemComp}
\end{align}
This property implies that a Codazzi tensor is also Weyl compatible, with the Weyl tensor $C_{jklm}$ replacing $R_{jklm}$. 
The contraction with the metric tensor $g^{il}$ gives $\mathscr C_{jm}R_k{}^m = \mathscr C_{km}R_j{}^m$, i.e. a Codazzi tensor
commutes with the Ricci tensor. 

As anticipated, we investigate two forms of Codazzi tensor. We name them in analogy with terms of an energy-momentum tensor: 
$\mathscr C_{jk}= Au_j u_k + Bg_{jk}$ (perfect fluid) and $\mathscr C_{jk}= \lambda (u_j \dot u_k + \dot u_j u_k)$ (current flow). 
$A\neq 0$, $B$, $\lambda $ are scalar fields. The vector field $u_j$ is time-like unit, $u^j u_j =-1$, and is named velocity. The vector field $\dot u_j =u^k\nabla_k u_j$ is spacelike, orthogonal to the velocity, and is named acceleration. \\
We show that the Codazzi property of such tensors strongly restricts the space-times they live in.
\subsection{}{\bf Perfect fluid Codazzi tensors and Stephani universes.}
\begin{thrm}\label{ONE} The perfect fluid tensor $\mathscr C_{jk}= Au_j u_k + Bg_{jk}$ with $u^ju_j=-1$ is Codazzi if and only if
\begin{align}
&\nabla_i u_j = \varphi (g_{ij}+u_i u_j) - u_i \dot u_j \label{C1}\\
&\nabla_i\varphi = - u_i \dot \varphi \label{C2}\\
&\nabla_i A = -u_i \dot A - \dot u_i A \label{C3}\\
&\nabla_i B = -u_i \dot B \label{C4}\\
&\varphi = -\dot B/A \label{C5}
\end{align}
\begin{proof}
See Appendix 1.
\end{proof}
\end{thrm}
%
%
\begin{prop}
If $\mathscr C_{jk}$ is a perfect-fluid Codazzi tensor, the velocity $u_i$ is Riemann compatible, 
$ u_i R_{jklm}u^m  + u_j R_{kilm}u^m  + u_k R_{ijlm}u^m =0 $, and
it is an eigenvector of the Ricci tensor, $R_{jk}u^k =\gamma u_j$, with eigenvalue
\begin{align}
\gamma = (n-1) (\dot\varphi +\varphi^2) - \nabla_k \dot u^k   \label{EIGENVALUE}
\end{align}
The following identity for the acceleration holds:
\begin{align}
 (\varphi \dot u_k  +\ddot u_k) u_l - u_k(\varphi \dot u_l + \ddot u_l)  = \nabla_k\dot u_l - \nabla_l\dot u_k. \label{SYMM}
 \end{align}
\begin{proof}
The first statement is an obvious consequence of \eqref{RiemComp} and of the first Bianchi identity. For the eigenvalue we evaluate:
\begin{align*}
R&_{jklm}u^m = \nabla_j \nabla_k u_l - \nabla_k \nabla_j u_l \\
=&\nabla_j [\varphi (g_{kl}+u_k u_l)-u_k\dot u_l] - \nabla_k [\varphi (g_{jl}+u_j u_l)-u_j\dot u_l]
\nonumber\\
=& - (g_{kl} u_j -g_{jl} u_k)\dot \varphi +(\nabla_j u_k-\nabla_k u_j) (\varphi u_l -\dot u_l) \\
&+ u_k ( \varphi \nabla_j u_l -\nabla_j \dot u_l) - u_j ( \varphi \nabla_k u_l -\nabla_k \dot u_l) \nonumber\\
=& - (g_{kl} u_j -g_{jl} u_k)(\dot \varphi +\varphi^2)- (u_j \dot u_k- u_k \dot u_j) (\varphi u_l -\dot u_l)
 - u_k \nabla_j \dot u_l + u_j \nabla_k \dot u_l
\end{align*}
The contraction with $g^{jl}$ gives: $R_{km}u^m = (n-1)(\dot \varphi +\varphi^2) u_k + \varphi \dot u_k
- u_k \eta - u_k \nabla_j \dot u^j + u^j \nabla_k \dot u_j $. 
Since $\dot u^j u_j=0$, the last term is: $-\dot u^j \nabla_k u_j = -\varphi \dot u_k +u_k \eta$
by eq.\eqref{C1}, and cancels three terms. The eigenvalue $\gamma $ is read.\\
The contraction with $u^j$ gives the symmetric tensor 
\begin{align}
u^jR_{jklm}u^m 
= (g_{kl} +u_l u_k)(\dot \varphi +\varphi^2) 
 + \dot u_k (\varphi u_l -\dot u_l) - u_k \ddot u_l - \nabla_k \dot u_l \label{RIEMUU}
\end{align} 
Subtraction with indices $k,l$ exchanged gives the identity for the acceleration.
\end{proof}
\end{prop}

 We the aid of the Weyl tensor, we obtain the expression of the Ricci tensor on a space-time with a perfect fluid Codazzi tensor.

 \begin{prop}[The Ricci tensor]\label{23}
 \begin{align}
R_{kl} =& \frac{R-n\gamma}{n-1} u_k u_l + \frac{R-\gamma}{n-1}g_{kl} +\Pi_{kl} \label{RICCITENSOR} \\
\Pi_{kl}=&\tfrac{1}{2} (n-2) [u_k(\varphi \dot u_l - \ddot u_l) + u_l (\varphi \dot u_k - \ddot u_k)
- (\nabla_k \dot u_l + \nabla_l \dot u_k)] \nonumber\\ 
&- (n-2) [\dot u_k \dot u_l +E_{kl} ] 
 + \frac{n-2}{n-1} (g_{kl}+u_k u_l) \nabla_p \dot u^p \nonumber
\end{align}
where $\gamma $ is the eigenvalue \eqref{EIGENVALUE}, $\Pi_{kl}$ is symmetric traceless and $\Pi_{kl} u^l=0$.
\begin{proof} 
The general expression of the Weyl tensor is:
\begin{align*}
C_{jklm} = R_{jklm} + \frac{ g_{jm} R_{kl} - g_{km} R_{jl} + g_{kl}R_{jm} -
g_{jl} R_{km}}{n-2} - R \frac{ g_{jm}g_{kl}-
g_{km}g_{jl} }{(n-1)(n-2)}
\end{align*}
$E_{kl}=u^j u^m C_{jklm}$ is the electric tensor. It is symmetric, traceless, with $E_{jk}u^k=0$. 
A double contraction and \eqref{RIEMUU} give:
\begin{align*}
E_{kl} = (g_{kl} +u_l u_k)(\dot \varphi +\varphi^2) + \dot u_k (\varphi u_l -\dot u_l) - u_k \ddot u_l - \nabla_k \dot u_l \\
- \frac{ R_{kl} +2 \gamma u_k u_j +\gamma  g_{kl} }{n-2} + R \frac{g_{kl}+u_ku_l}{(n-1)(n-2)}.
\end{align*}
The Ricci tensor is obtained:
\begin{align*}
R_{kl} =& \left [ \frac{R-n\gamma +(n-2)\nabla_p\dot u^p}{n-1}  \right ] u_k u_l 
+ \left [ \frac{R-\gamma +(n-2)\nabla_p\dot u^p}{n-1}  \right ] g_{kl}\\
& - (n-2)[ \dot u_k \dot u_l -\varphi \dot u_k u_l + u_k\ddot u_l +\nabla_k \dot u_l +E_{kl}].
\end{align*}
The expression is symmetrized with the identity \eqref{SYMM} and the correction to the perfect fluid part is made traceless by
subtraction. 
\end{proof}
\end{prop}
%
%
We discuss the geometric restrictions posed by a perfect-fluid Codazzi tensor. 
The presence of a shear-free and vorticity-free velocity field, eq.\eqref{C1}, classifies the space-time as {\em doubly-twisted} \cite{MM21}, 
i.e. there is a coordinate frame such that the metric has the form \eqref{DTW}.\\
 In this frame, with the Christoffel symbols
$$ \Gamma_{00}^0 =\frac{\partial_t b}{b}, \quad \Gamma_{\mu0}^0 =\frac{\partial_\mu b}{b}, \quad \Gamma_{\mu\nu}^0 = \frac{\partial_t a}{ab^2}
g^\star_{\mu\nu}, \quad \Gamma_{0\mu}^\nu = \frac{\partial_t a}{a}\delta_\mu^\nu$$
eq.\eqref{C1} for $u_j$ and  $\dot u_j=u^k\nabla_k u_j$ give: $u_0=-b(t,{\bf x})$, $u_\mu =0$, and
\begin{align}
 \dot u_0=0, \; \dot u_\mu = \frac{\partial_\mu b(t,{\bf x})}{b(t,{\bf x})};\quad 
\varphi = \frac{1}{b(t,{\bf x})} \frac{\partial_t a(t,{\bf x})}{a(t,\bf x)} .
\end{align}
By eq.\eqref{C2}, the doubly twisted metric has the constraint that $\varphi $ only depends on time. With
$a=1/V({\bf x},t)$, the metric \eqref{DTW} with the constraint becomes:
\begin{align}
ds^2 = -\left[ \frac{1}{\varphi (t)} \frac{\partial_t V}{V}\right ]^2 dt^2 + \frac{g_{\mu\nu}^\star ({\bf x})dx^\mu dx^\nu
}{V^2({\bf x},t)}
\end{align}
This metric generalizes the well known Stephani metrics, presented in the following example.

\begin{ex}
Remarkably, equations \eqref{C1}--\eqref{C5} coincide with eqs. 37.32--37.34 in the book by Stephani et al. \cite{Stephani}. They were derived for a Riemann tensor of the form $R_{jklm}=\mathscr C_{jl}\mathscr C_{km}-\mathscr C_{jm}\mathscr C_{kl}$, with $\mathscr C_{jk}=Au_ku_l+
Bg_{jk}$ (note that if $\mathscr C_{jk}$ is invertible then, the Bianchi identity implies that it is a Codazzi tensor \cite{Goenner}). 
Such space-times are conformally flat and are named Stephani universes \cite{Stephani}\cite{Krasinski}.  
They are solutions of the Einstein equation with a perfect fluid source $T_{jk}$.\\
The Stephani metric in $n=4$ is
$$ds^2 =  -\left[ \frac{1}{\varphi (t)} \frac{\partial_t V}{V}\right ]^2 dt^2 + \frac{dx^2+dy^2+dz^2}{V^2({\bf x},t)} $$
with $V ({\bf x},t)= V_0(t) +\frac{B^2(t) -\varphi^2(t)}{4V_0(t)} \| {\bf x} - {\bf x}_0(t)\|^2$, where $V_0$, $\varphi $ and ${\bf x}_0$ 
are arbitrary functions of time.
\end{ex}

We now consider some special conditions of the perfect fluid Codazzi tensor.

\begin{lem} If the acceleration is closed, $\nabla_j \dot u_k =\nabla_k \dot u_j$, then $b(t,{\bf x}) = \hat b(t) b({\bf x})$, and
$\ddot u_k = \eta u_k - \varphi \dot u_k $.
\begin{proof} The condition that matters is $\nabla_0 \dot u_\mu = \nabla_\mu \dot u_0$ i.e. $\partial_t \dot u_\mu - \Gamma_{0\mu}^\nu \dot u_\nu = 
-\Gamma_{\mu 0}^\nu \dot u_\nu$. By the symmetry of the Christoffel symbols, we remain with $0=\partial_t \dot u_\mu$ i.e.  $\dot u_\mu =\partial_\mu \log b$ is independent of $t$. Then $b(t,{\bf x}) = b_1(t)b_2({\bf x})$.\\
Eq.\eqref{SYMM} now is: $ (\varphi \dot u_k  +\ddot u_k) u_l - u_k(\varphi \dot u_l + \ddot u_l) =0$. Contraction 
with $u^l$ is: $\ddot u_k = - \varphi \dot u_k - u_k (u^l\ddot u_l)$. The identity $u^l\dot u_l=0$ gives $u^l\ddot u_l = -
\dot u^l \dot u_l \equiv -\eta$.
\end{proof}
\end{lem}

\noindent
$\bullet $ If $\nabla_k A=-u_k \dot A $ i.e. $\dot u_k=0$, then $b(t,{\bf x})$ is only a function of time. It is $b=1$ after a rescaling of time. The equations
$\partial_\mu \varphi =0$ show that $a$ only depends on time. Therefore, the space-time is a {\em generalised Robertson Walker} (GRW) space-time, 
eq.\eqref{GRW} \cite{Chen2014,MMGRW}.\\
This agrees with Theorem 1.2 in \cite{Merton13},  stating that (in a Riemannian setting) a perfect fluid Codazzi tensor such that 
$h^{jk}\nabla_k \mathscr C^i{}_i =0$ implies a warped metric.\\ 
With $\xi \equiv (n-1)(\dot\varphi +\varphi^2)$, the Ricci tensor now is:
\begin{align}
R_{jk} = \frac{R-n\xi}{n-1} u_j u_k + \frac{R-\xi}{n-1} g_{jk} -(n-2) E_{jk}. \label{RICCIT}
\end{align}
\noindent
$\bullet $ If $B=0$, i.e. $\mathscr C_{jk}=A u_j u_k$, then $\nabla_i u_j = -u_i\dot u_j$ and $A$ solves \eqref{C3}. The equation $\varphi =0$
gives that $a(t,{\bf x})$ is independent of time, and can be absorbed in the space metric to give 
$$ds^2 = -b^2(t,{\bf x}) dt^2 +g^\star_{\mu\nu}({\bf x}) dx^\mu dx^\nu $$
Its conformally flat and spherically symmetric version generalises the Schwarzschild interior solution, eq.37.39 in \cite{Stephani}.
If moreover $\dot u_i$ is closed, then the metric is {\em static} (\cite{Stephani}, page 283):
\begin{align}
 ds^2 = - b^2({\bf x}) dt^2 +  g_{\mu\nu}^\star ({\bf x}) dx^\mu dx^\nu. \label{STAT}
\end{align}
$\bullet $ In General Relativity the vanishing of the Cotton tensor ${\sf C}_{jkl}=0$ means that $R_{kl}-g_{kl} \frac{R}{2(n-1)}$ is a Codazzi tensor. The Einstein equations then imply that also 
$\mathscr C_{kl}= T_{kl}-\frac{T}{n-1}g_{kl}$ is a Codazzi tensor.

\subsection{}{\bf Current-flow Codazzi tensors}\quad\\
We investigate Codazzi tensors with the form of a current-flow tensor $\mathscr C_{jk} =\lambda (u_j \dot u_k + \dot u_j u_k),$ with closed $\dot u_i$.

The eigenvalues are $0$ and $\pm i\lambda \sqrt\eta $, the latter being non-degenerate with complex eigenvectors 
$V^{\pm}_k =\pm \sqrt \eta u_k + i\dot u_k$, $g^{jk}V^+_j V^-_k =0$. 
Since the Codazzi tensor commutes with the Ricci tensor, $V^{\pm}_k$ are also eigenvectors of the Ricci tensor. From $0=V^+_j R^{jk}V^-_k$ 
one obtains 
\begin{align}
\dot u^j R_{jk}\dot u^k = -\eta u^j R_{jk} u^k \label{RICCIUU}
\end{align}

\begin{thrm}\label{Cflow}
The tensor $\mathscr C_{jk} =\lambda (u_j \dot u_k + \dot u_j u_k)$ with closed acceleration is Codazzi if and only if:
\begin{align}
&\nabla_j u_k = -\frac{\dot \lambda}{\lambda}\frac{ \dot u_j \dot u_k }{\eta}- u_j \dot u_k \label{D1}\\
&\nabla_j \lambda = -u_j \dot \lambda -\lambda\dot u_j \left( 2+\frac{\dot u^p\nabla_p \eta}{2\eta^2}\right)  \label{D2}\\
&\nabla_j \dot u_k = - \eta u_j u_k -\frac{\dot\lambda}{\lambda} (\dot u_j u_k + u_j \dot u_k) \label{D3} 
+\dot u_j \dot u_k \frac{\dot u^p\nabla_p\eta}{2\eta^2}
\end{align}
\begin{proof}
See Appendix 2.
\end{proof}
\end{thrm} 
A useful relation found in the proof is
\begin{align}
\nabla_k \eta = -2 \frac{\dot\lambda}{\lambda}\eta u_k +  \dot u_k  \frac{\dot u^p \nabla_p \eta}{\eta}. \label{nablaeta}
\end{align}

We discuss the geometric restrictions posed by a current-flow Codazzi tensor with closed acceleration. \\
Since the velocity has non-zero shear tensor
$$\sigma_{jk} =  \frac{\dot \lambda}{\lambda}\left[ \frac{g_{jk}+u_j u_k}{n-1}- \frac{\dot u_j \dot u_k}{\eta}\right ] $$
there are coordinates such that the metric has the structure \cite{Ferrando}:
\begin{align}
ds^2 = - b^2(t,{\bf x}) + G^\star_{\mu\nu} (t,{\bf x})dx^\mu dx^\nu \label{FERR}
\end{align}
with Christoffel symbols $\Gamma_{00}^0 =\frac{\partial_t b}{b}$, $\Gamma_{\mu0}^0 =\frac{\partial_\mu b}{b}$, 
$\Gamma^\mu_{00} = G^{\star \mu\nu} b\partial_\nu b$, 
$ \Gamma_{\mu\nu}^0 = \frac{\partial_t G^{\star}_{\mu\nu}}{2b^2}$, $\Gamma_{0\nu}^\mu = \tfrac{1}{2}G^{\star \mu\rho} \partial_t G^\star_{\nu\rho}$ and $\Gamma_{\rho\sigma}^\mu = \Gamma_{\rho\sigma}^{\star\mu}$. 
The equations for $u$, $\dot u$  give:  
$$u_0=-b(t,{\bf x}),\; u_\mu =0, \qquad \dot u_0=0, \; \dot u_\mu = \frac{\partial_\mu b(t,{\bf x})}{b(t,{\bf x})} $$
In this frame, the equations $\nabla_\mu u_\nu = -\frac{\dot\lambda}{\lambda} \frac{\dot u_\mu \dot u_\nu}{\eta} $ and $\nabla_0\dot u_\mu = 
-\frac{\dot\lambda}{\lambda}u_0\dot u_\mu$ are:
\begin{align}
- \tfrac{1}{2} b \,\frac{\partial G^{\star}_{\mu\nu}}{\partial t} =\frac{\dot \lambda}{\lambda} \frac{\partial_\mu b \,\partial_\nu b}{\eta} , \quad
\frac{\partial \dot u_\mu}{\partial t}  -\tfrac{1}{2} \dot u_\nu G^{\star \nu\rho}\frac{\partial G^\star_{\mu\rho} }{\partial t} = \frac{\dot\lambda}{\lambda}b\, \dot u_\mu \label{Gmunu}
\end{align}
We now specialize to static space-times.

\subsubsection{Static space-times}  
If $\dot \lambda =0$, 
eq.\eqref{Gmunu} shows that $G^\star_{\mu\nu} $ is independent of time $t$, as well as  $\dot u_\mu$. 
Then $b(t,{\bf x}) =\beta (t) b({\bf x})$. The product $\beta^2(t)dt^2$ in $ds^2$ redefines 
the time, and the metric is {\em static}, eq.\eqref{STAT}.\\
Theorem \ref{Cflow} becomes: the current-flow tensor with $\dot\lambda =0$
and closed acceleration is Codazzi if and only if:
\begin{align}
&\nabla_j u_k = - u_j \dot u_k \label{S1}\\
&\nabla_j \lambda = -\lambda\dot u_j \left( 2+\frac{\dot u^p\nabla_p \eta}{2\eta^2}\right)  \label{S2}\\
&\nabla_j \dot u_k = - \eta u_j u_k +\dot u_j \dot u_k \frac{\dot u^p\nabla_p\eta}{2\eta^2} \label{S3}
\end{align}
Eq.\eqref{S1} and closedness of $\dot u_i$ covariantly confirm the space-time as static. 

\begin{prop}\label{PPP}
In a static space-time eqs.\eqref{S1}-\eqref{S3} with closed $\dot u_i$, the vectors $u_i$ and $\dot u_i$ are eigenvectors of
the Ricci tensor with the same eigenvalue.
\begin{proof}
1) For brevity, put $\epsilon =\dot u^p\nabla_p \eta$.\\ Eq.\eqref{nablaeta} with $\dot\lambda =0$ is: $\nabla_j \eta^2 = 2 \epsilon \dot u_j$
Now it is $\nabla_k \nabla_j \eta^2 = 2\dot u_j \nabla_k \epsilon + 2 \epsilon \nabla_k \dot u_j$. Antisimmetrization, with the property that
$\nabla_k\dot u_j = \nabla_j \dot u_k$ gives: $\dot u_j \nabla_k \epsilon  = \dot u_k \nabla_j \epsilon $ i.e.
$$ \nabla_j \epsilon = \dot u_j \frac{\dot u^p\nabla_p \epsilon}{\eta} $$
2) $R_{jklm}u^m = \nabla_j \nabla_k u_l - \nabla_k \nabla_j u_l  = -\nabla_j (u_k\dot u_l)+\nabla_k (u_j\dot u_l) = (u_j \dot u_k - u_k\dot u_j)\dot u_l -
u_k \nabla_j\dot u_l + u_j \nabla_k \dot u_l = (u_j \dot u_k - u_k\dot u_j)\dot u_l (1 + \epsilon/2\eta^2)$. Contraction with $g^{jl}$:
\begin{align}
 R_{km} u^m = - (\eta + \frac{\epsilon}{2\eta} ) u_k \label{RU}
 \end{align}
3) $R_{jklm}\dot u^m = \nabla_j \nabla_k \dot u_l - \nabla_k \nabla_j \dot u_l  = \nabla_j (-\eta u_k u_l + \dot u_k \dot u_l \frac{\epsilon}{2\eta^2}) -
\nabla_k (-\eta u_j u_l + \dot u_j \dot u_l \frac{\epsilon}{2\eta^2})= -(u_k \nabla_j\eta - u_j\nabla_k\eta) u_l +\eta (u_j \dot u_k - \dot u_j u_k)u_l +
(\dot u_k \nabla_j \dot u_l - \dot u_j \nabla_k \dot u_l)\frac{\epsilon}{2\eta^2} +(\dot u_k\nabla_j \frac{\epsilon}{2\eta^2} - \dot u_j
\nabla_k \frac{\epsilon}{2\eta^2})\dot u_l $. The last parenthesis is zero because $\nabla_j \frac{\epsilon}{2\eta^2}$ is proportional to $\dot u_j$.
Then: $R_{jklm}\dot u^m =  (u_j \dot u_k - \dot u_j u_k)u_l (\eta + \frac{\epsilon}{\eta}) + (\dot u_k \nabla_j \dot u_l - 
\dot u_j \nabla_k \dot u_l)\frac{\epsilon}{2\eta^2}$. Contraction with $g^{jl}$:
\begin{align}
 R_{km} \dot u^m =  - \left (\eta +\frac{\epsilon}{2\eta}   \right)  \dot u_k \label{RDOTU}
\end{align}
\end{proof}
\end{prop}
The Ricci tensor is now obtained.\\ 
In Prop.\ref{PPP} we evaluated 
$R_{jklm}u^m = (u_j \dot u_k - u_k\dot u_j)\dot u_l (1 + \epsilon/2\eta^2)$. Contraction
with $u^j$ is $u^j R_{jklm}u^m = - \dot u_k \dot u_l (1 + \epsilon/2\eta^2)$. 
The contraction of the Weyl tensor and \eqref{RU} give 
\begin{align*}
E_{kl} =  -\dot u_k \dot u_l (1 + \frac{\epsilon}{2\eta^2})- \frac{ R_{kl} - (g_{kl}+2u_k u_l) (\eta + \epsilon/2\eta)}{n-2}
 + R \frac{g_{kl}+u_k u_l}{(n-1)(n-2)}
\end{align*}
We then find:
\begin{align}
R_{kl}= \left[ \frac{R}{n-1}+ 2\eta +\frac{\epsilon}{\eta}\right ] u_k u_l +\left[ \frac{R}{n-1}+ \eta +\frac{\epsilon}{2\eta}\right ]g_{kl}  
 -(n-2) \left[E_{kl} +\dot u_k \dot u_l (1 + \frac{\epsilon}{2\eta^2})\right ].\label{RICCIST2}
\end{align}
In particular, by eq.\eqref{RDOTU}, one has the eigenvalue equation
$$(n-2)E_{kl} \dot u^l = \left[\frac{R}{n-1}-(n-4)(\eta+\frac{\epsilon}{2\eta})\right ] \dot u_k$$
Eq.\eqref{S3} with $\epsilon =0$ (then $\eta $ is a constant) was obtained by Rao and Rao \cite{RaoRao} 
in a static metric to characterize the relativistic generalisation of the uniform Newton force at a spatial
hypersurface.

\subsubsection{} 
We restrict the static space-time to be {\em spherically symmetric}, and give some examples in the end:
\begin{align}
ds^2 = -b^2(r) dt^2 + f_1^2(r) dr^2 + f_2^2(r) d\Omega_{n-2}^2
\end{align} 
In spherical symmetry $\dot u$ is radial and $\dot u_r = b'(r)/b(r)$ (a prime is a derivative in $r$).
The definition $\eta = \dot u^k \dot u_k$ gives:
\begin{align}
\eta (r)=  \frac{1}{f_1^2(r)} \frac{b'^2(r)}{b^2(r)} \label{f1f1}
 \end{align}
In such coordinates the solution of eq.\eqref{S2} is
\begin{align}
\lambda (r) = \kappa\frac{f_1(r)}{b(r)b'(r)} \label{LALA}
\end{align}
with a constant $\kappa $.\\
Since $\dot u$ is a radial vector, the angular components of eq.\eqref{S3} are $\nabla_a \dot u_{a'} =0$ (where $a, a' = 1,...,n-2$ enumerate the angles). It implies $\Gamma_{a,a'}^r \dot u_r= 0$ i.e. $\Gamma_{a,a'}^r = 0$. With the expression in \cite{MMDW} Appendix 9.6, one gets the condition on the metric:
$$ \frac{df_2}{dr}=0. $$
In conclusion, a static spherically symmetric space-time with Codazzi tensor $\mathscr C_{jk}=\lambda (u_j\dot u_k + \dot u_j u_k)$ 
with closed acceleration has the form:
\begin{align}
ds^2 = - b^2(r) dt^2 + f_1^2(r) dr^2 + L^2 d\Omega_{n-2}^2  \label{METRICL}
\end{align}
where $L$ is a positive constant. \\
The electric tensor and the scalar curvature of the space manifold are obtained from eq.(33) in \cite{MMDW} with $a=1$, $f_2=L$, and the relations \eqref{f1f1} and \eqref{LALA}:
\begin{align}
&E_{jk}(r)  = \frac{n-3}{n-2}\frac{1}{f_1^2} \left( \frac{f_1^2}{L^2} +\frac{b'}{b} \frac{f_1'}{f_1} - \frac{b''}{b} \right )\ \left [ \frac{\dot u_j \dot u_k}{\eta} - \frac{h_{jk}}{n-1}\right ] \label{EEE}\\
&R^\star = \frac{1}{L^2}(n-2)(n-3)
 \end{align}
 where $h_{jk} = g_{jk}+ u_k u_l$.\\
 The Ricci tensor \eqref{RICCIST2} in spherical coordinates is sum of three tensors, proportional to $u_j u_k$, $g_{jk}$ and $\dot u_j \dot u_k$. \\

We list some examples of the metric \eqref{METRICL}. They share the same form of Ricci tensor \eqref{RICCIST2}, with electric tensor \eqref{EEE}. Moreover,
they are endowed with a current-flow Codazzi tensor with non-zero components $\mathscr C_{0r}=\mathscr C_{r0} =\kappa f_1(r)/b(r)$
in the coordinates of each below-listed metric.

\begin{ex}\label{E1} {\sf Nariai space-times} solve the Einstein equations in vacuum \cite{Nariai51,Bousso,Ortaggio,Cardoso04}:
$$ ds^2= \frac{1}{\Lambda} \left[ -a\cos\log (\frac{r}{r_0} )dt^2 + \frac{1}{r^2} (dr^2 + r^2 d\Omega_2^2)\right ]. $$
\end{ex}
\begin{ex}\label{E2} {\sf Bertotti - Robinson space-times} are conformally flat solutions of the source-free Einstein-Maxwell equations
with non-null e.m. field \cite{Bertotti,Robinson}:
\begin{align} ds^2 = \frac{r_0^2}{r^2}\left [ -dt^2 + dr^2 +r^2 d\Omega_{n-2}^2 \right]. \label{BR}
\end{align}
The Ricci tensor is \eqref{RICCIST2} with $R=0$, $\epsilon=0$, $\eta=\frac{1}{r_0^2}$ $\lambda = -\kappa \frac{r^2}{r_0^2}$:
\begin{align}
 R_{kl}=\frac{1}{r_0^2}(2 u_k u_l +g_{kl})-2 \dot u_k \dot u_l.
 \end{align}
Eqs.\eqref{S1} and \eqref{S3}, that reads $\nabla_j \dot u_k =\frac{1}{r_0^2} u_j u_k$, imply that 
$\nabla_i R_{jk}=\nabla_j R_{ik}$. Therefore, Bertotti-Robinson space-times have two Codazzi tensors:
the Ricci tensor and  $\mathscr C_{jk}=-\kappa \frac{r^2}{r_0^2}(u_j \dot u_k + \dot u_j u_k)$.
\end{ex}
\begin{ex}\label{E3} 
In \cite{Gurses} black holes are studied in string-corrected Einstein-Maxwell theory coupled to a dilaton field. The solution displayed in eq.35 is
$$ ds^2 = -(ar^2+br +c) dt^2 + \frac{dr^2}{ar^2+br +c} + L^2 d\Omega_2^2. $$
\end{ex}
\begin{ex}\label{E4} In \cite{Lowe} the Bertotti-Robinson-type black hole solutions of string theory are obtained, by CFT methods. This one (eq. 38) is an example:
\begin{align*}
 ds^2 = -\left[ \tfrac{r^2}{\ell^2} + \tfrac{J^2}{r^2}-M\right] dt^2 + \frac{dr^2}{\left[ \tfrac{r^2}{\ell^2} + \tfrac{J^2}{r^2}-M\right]}
 + L^2 d\Omega^2_2 
\end{align*}
where $M$ is the mass, $J$ is the angular momentum, $\ell^2$ is proportional to the cosmological constant.
\end{ex}
\begin{ex}\label{E5} In \cite{Ast} spherical black hole solutions of the Einstein-Maxwell-scalar equations are found, where the scalar field is non-minimally coupled to the Maxwell invariant. Among others the following metric is given (eq.4.11), where
$a$ is a constant: 
\begin{align*}
ds^2 = a \left ( -r^2 dt^2 +\frac{1}{r^2}dr^2 \right ) + L^2 \,d\Omega_2^2. 
\end{align*}
\end{ex}
\quad

\subsection{}{\bf Yang Pure space-times}\quad\\
A {\sf Yang Pure space-time} is defined by a Ricci tensor that is a Codazzi tensor:
\begin{align}
\nabla_j R_{kl} = \nabla_k R_{jl} \label{Yang}
\end{align}
equivalent to $\nabla^m R_{jklm}=0$. Contraction with $g^{jl}$ gives $\nabla_k R=0$. 
They were introduced by Chen Ning Yang in 1974 in the geometry of Yang-Mills theories  \cite{Yang}.\\
These are examples of solutions of Yang's equation \eqref{Yang}.\\
$\bullet$ Vacuum solutions of Einstein's equations: $R_{kl}=\Lambda g_{kl}$.\\
$\bullet$ Wei-Tou Ni obtained the conformally-flat non-static solution \cite{WTNi}
$$ ds^2 = \left[C+\frac{f(r-t)}{r}+\frac{g(r+t)}{r}\right ] (-dt^2 +dr^2 +r^2 d\Omega_2^2)$$
where $C$ is a constant, $f$ and $g$ are arbitrary functions, and also the solution:
$$ ds^2 = -dt^2 + \left [1 + \frac{a}{r} + b r^2 \right ]^{-1} dr^2 + r^2 d\Omega_2^2$$
$\bullet$ In 1975 A. H. Thompson \cite{Thompson} found geometrically degenerate solutions of Yang’s gravitational
equations. In particular, he showed that the Bertotti-Robinson metric eq.\eqref{BR} is Yang Pure.\\
$\bullet$ Friedmann-Robertson-Walker (FRW) space-times 
$$ ds^2 = -dt^2 + a^2(t)\left[\frac{dr^2}{1- k r^2} + r^2 d\Omega_2^2 \right ] $$
may be also characterized by a ``perfect fluid'' Ricci tensor
$$ R_{jk} = \tfrac{1}{3}(R-4\xi) u_j u_k + \tfrac{1}{3} (R-\xi) g_{jk} $$
and zero Weyl tensor. Here: $u^k u_k=-1$, $\nabla_i u_j = H (u_iu_j + g_{ij})$ where $H=\dot a/a$ is Hubble's constant and $\xi = 3(H^2+\dot H) = 3\ddot a/a$.
The Cotton tensor being zero, a FRW space-time is Yang Pure if and only if $\nabla_j R=0$.\\ 
The flat case $k=0$ was solved by the authors \cite{MMRW}. While the two geometric constraints fix the Ricci tensor, 
the Einstein equations provide a source which is a perfect fluid with equation of state $p=w(t)\rho $ that evolves from $w=1/3$ (pure radiation) to $w=-1$ (accelerated expansion, without a cosmological constant $\Lambda$).

\section{Harada-Cotton gravity}
The results of the previous section are interesting for Harada's Cotton gravity. 
The symmetries of the Weyl tensor imply two important facts \cite{Harada21}:
\begin{enumerate}
\item $g^{kl}{\sf C}_{jkl}=0$, then \eqref{HEq} mantains the  law $0=\nabla_k T^{jk}$.
\item {$\nabla^l {\sf C}_{jkl}=0$ implies that $R_{jk}$ and $T_{jk}$ commute:\\
$0 =  \nabla_l (\nabla_j T_k{}^l-  \nabla_k T_j{}^l) \\
= [\nabla_l ,\nabla_j ]T_k{}^l -[\nabla_l,\nabla_k]T_j{}^l +  \nabla_j (\nabla_l T_k{}^l) - \nabla_k (\nabla_l T_j{}^l)\\
=  R_{ljkm}T^{ml}+R_{jm} T_k{}^m - R_{lkjm}T^{ml} - R_{km} T_j{}^m$.\\
The first term cancels the third one. }
\end{enumerate}
As stated in the introduction, eq.\eqref{HEq} naturally provides the Codazzi tensor in eq.\eqref{CRT}.
Depending on its form, there are different levels of Cotton gravity, that are
extensions of the Einstein gravity. \\
The choice of the Codazzi tensor restricts the space-time which, in turn, provides the structure of the Ricci tensor. Together, the Ricci and the Codazzi tensors determine the energy-momentum tensor:
\begin{align}
T_{kl} = R_{kl} - \frac{1}{2} R g_{kl}  -\mathscr C_{kl} +g_{kl}\mathscr C^j{}_j\label{ENMOM}
\end{align}
By construction, the metric of the space-time solves the Cotton-gravity equation \eqref{HEq} with the energy-momentum tensor \eqref{ENMOM}.
This approach reverses the standard one, with the matter tensor as input. \\
Eq.\eqref{ENMOM} is Einstein's equation corrected by a Codazzi tensor, eq.\eqref{EINSTEIN+COD}, in analogy with other theories of extended gravity (the $H$-term of eq.26 in \cite{Cap15}).

\subsection{}{\bf Yang Pure spaces.}\quad\\
Since the Ricci tensor is Codazzi, the definition \eqref{Cott}  of Cotton tensor
shows that Yang Pure spaces are solutions of the vacuum Harada equations ${\sf C}_{jkl}=0$.\\

Now we present the simplest Codazzi tensors, with examples that only aim at illustrating the procedure.\\

\subsection{} The trivial Codazzi tensors  $\mathscr C_{jk}=0$ and $\mathscr C_{jk} = B g_{jk}$ (with $B $ constant by the Codazzi condition) 
give the Einstein equations without or with a cosmological constant.

\subsection{Case $\mathscr C_{jk} \boldsymbol{= A u_j u_k+ Bg_{jk}}$, $\boldsymbol{u^ku_k=-1}$}\quad\\
The generalized Stephani Universes are solutions of the Harada equation with energy-momentum tensor \eqref{ENMOM} built with
the Ricci tensor \eqref{RICCITENSOR} and the Codazzi tensor.\\ 
Such inhomogeneous cosmological models may provide an explanation of the observed accelerated expansion of the universe and bypass the dark energy problem
(see for example \cite{Dabrowski,Balcerzak} and references therein).\\
We here give the Ricci tensor for the simpler Stephani Universe in $n=4$:
\begin{align}
R_{kl} = 2AB \, u_k u_l +g_{kl} (3B^2 -AB) \label{STPH}
\end{align} 
Its perfect fluid form implies a perfect fluid source in the Einstein equations, as well as in the Harada equations
(with different density and pressure).

\subsection{\bf Case $\mathscr C_{jk} = \boldsymbol{A u_j u_k}$, $\boldsymbol{u_k u^k=-1}$, $\boldsymbol{\dot u}$ closed}\quad\\ 
The Codazzi condition is equivalent to $\nabla_i u_j = -u_i\dot u_j$
and \eqref{C3}. The space-time is static and the velocity is eigenvector of the Ricci tensor: $R_{jk}u^k=-u_j (\nabla_k \dot u^k)$.\\
This example in $n=4$ is static and spherically symmetric:
 \begin{align}
 ds^2 = -b^2(r) dt^2 + f(r)^2 dr^2 + r^2 d\Omega_2^2 
 \end{align}
The function $A(r)$ solves \eqref{C3}, where the time component is an identity and $A' = -A b'/b $ (a prime is a derivative in $r$). The equation 
is solved by 
$$A(r) = \frac{k}{b(r)}$$ 
where $k$ is a constant. The covariant form of the Ricci tensor on static isotropic space-times was obtained in \cite{MMDW} (eq.49 with $\varphi =0$):
\begin{align}
&R_{jk} = u_j u_k \frac{R+4\nabla_p \dot u^p}{3}+ g_{jk}\frac{R  + \nabla_p \dot u^p}{3} + \Pi_{jk} \label{RICCIST}\\
& \Pi_{jk}=\left [\frac{\dot u_j \dot u_k}{\eta} -\frac{h_{jk}}{3}\right ] \left [\nabla_p \dot u^p - 3 \left ( \eta + \frac{\dot u^i\nabla_i\eta}{2\eta} \right ) - 2 E(r)  \right ] \nonumber
\end{align} 
where $\eta = \dot u^j \dot u_j = b'^2/(b^2 f^2)$. $E(r)$ is the amplitude of the electric tensor
\begin{align}
 E(r)= \frac{1}{2f^2}\left[ \frac{f^2}{r^2} - \frac{1}{r^2}  - \frac{f'}{f r}  +\frac{b'}{b}\frac{d}{dr} \log (fr) - \frac{b''}{b}  \right ]  \label{E(r)}
 \end{align}
 \begin{align*}
&R=\frac{2}{r^2}\left (1-\frac{1}{f^2}\right )  + \frac{4}{r}
  \frac{f'}{f^3}  -\frac{2}{bf^2} \left (b'' - b' \frac{f'}{f} + 2\frac{b'}{r}\right )\\
& \nabla_p \dot u^p = \frac{1}{bf^2} \left (b'' - b' \frac{f'}{f} + 2\frac{b'}{r} \right )
\end{align*}
The traceless tensor $\Pi_{jk}$ modifies the perfect fluid term. It is $\Pi_{jk}u^k=0$ and $\Pi_{jk}\dot u^k \propto \dot u_j$. \\
The Ricci tensor has three eigenvalues and builds a Cotton tensor ${\sf C}_{jkl}$ that, by construction,
solves Harada's equation \eqref{HEq} for the following energy-momentum tensor:
 \begin{align*}
T_{jk} = &u_j u_k \frac{R+4\nabla_p \dot u^p}{3}+ g_{jk}\frac{R  + \nabla_p \dot u^p}{3} +g_{kl} \frac{T}{3} - g_{kl} \frac{R}{6} 
\\ &-\mathscr C_{kl}
 +\left [\frac{\dot u_j \dot u_k}{\eta} -\frac{h_{jk}}{3}\right ] \left [\nabla_p \dot u^p - 3 \left ( \eta + \frac{\dot u^i\nabla_i\eta}{2\eta} \right ) - 2 E(r) \right ].
\end{align*}
A simplification is done with the expression of the trace $T$, and with the following identity (Lemma 3.4 in \cite{MMDW}):
 \begin{align*}
\nabla_p \dot u^p - 3\left( \eta + \frac{\dot u^i\nabla_i \eta}{2\eta}  \right )  = 
- \frac{2}{f^2}\,\left [\frac{b''}{b}  - \frac{b'}{b} \frac{d}{dr} \log (r f)  \right ].
 \end{align*} 
 The result is:
\begin{align*}
T_{jk} =& u_j u_k \left [\frac{R}{3}+\frac{4}{3}\nabla_p \dot u^p - \frac{k}{b}\right] + g_{jk} \left[ -\frac{R}{6}  + \frac{1}{3}\nabla_p \dot u^p   -\frac{k}{b} \right ]\\
 &+\left [\frac{\dot u_j \dot u_k}{\eta} -\frac{h_{jk}}{3}\right ] \frac{1}{f^2} \left [ - \frac{b''}{b}  + \frac{b'}{b} \left(\frac{1}{r}+\frac{f'}{f}\right )  -  \frac{f^2-1}{r^2}   + \frac{f'}{f r}  \right ].
 \end{align*}
The tensor specifies the parameters of a static fluid
\begin{align*}
 T_{jk}= (P+\mu) u_j u_k + Pg_{jk}  + \left [\frac{\dot u_j \dot u_k}{\eta} -\frac{h_{jk}}{3}\right ] (p_r-p_\perp )
\end{align*}
with $P=\frac{1}{3}p_r + \frac{2}{3}p_\perp$ (effective pressure), density $\mu $, radial pressure $p_r$, transverse pressure $p_\perp$,
constructed with the free parameters $b(r)$, $f(r)$, $k$. Note the pressure anisotropy despite the spherical symmetry of the metric.

\subsection{Case $\mathscr C_{jk}\boldsymbol{=\lambda (u_j\dot u_k + \dot u_j u_k)}$ with closed $\boldsymbol{\dot u_j}$} \quad\\
The metrics in examples \ref{E1}-\ref{E5} are static spherically symmetric solutions of equations of various gravity
theories, Einstein, Einstein-Maxwell, low energy string, with their own matter or radiation content. However, since they all contain a current-flow Codazzi tensor, they all solve the Harada equation \eqref{HEq} with a proper energy-momentum tensor that is obtained below, characterized by a current-flow term. \\  
The metrics determine the Ricci tensor \eqref{RICCIST2} with radial symmetry. The energy-momentum tensor is \eqref{ENMOM} with $R=R^\star -2\eta - \epsilon/\eta$ and $R^\star =2/L^2$:
\begin{align*}
T_{kl}=&  u_k u_l  \frac{R^\star}{2}+ h_{kl}\left [ -\frac{R^\star}{6}+\frac{2\eta}{3}+\frac{\epsilon}{3\eta}\right]\\
&-\mathscr C_{kl}  - 2\left[ \frac{\dot u_k \dot u_l}{\eta }- \frac{h_{kl}}{3}\right] \left[ \eta +\frac{\epsilon}{2\eta} +E(r)\right]\\
E(r)=&\frac{1}{2}\frac{1}{f_1^2} \left( \frac{f_1^2}{L^2} +\frac{b'}{b} \frac{f_1'}{f_1} - \frac{b''}{b} \right ).
\end{align*}
It is the energy-momentum tensor of a fluid with velocity $u_j$, acceleration $\dot u_j$, energy density $\mu=\frac{1}{2}R^\star $, pressure anisotropy $p_r -p_\perp= -2\eta -\epsilon/\eta -2E(r)$ and effective pressure $3P=p_r +2p_\perp =
 -\frac{R^\star}{2}+2\eta+\frac{\epsilon}{\eta}$. 
 
 \begin{ex}
Consider the Bertotti-Robinson metric in \ref{E2}: $b(r)=f_1(r)= r_0/r$, $L=r_0$. It is
$\eta =1/r_0^2$, $\epsilon =0$ and $\lambda =-\kappa r^2/r_0$. \\
The Ricci tensor for this metric is:
$ R_{kl} = \frac{1}{r_0^2}(2u_k u_l  + g_{kl} ) - 2\dot u_k \dot u_l $ and $R=0$. The metric \eqref{BR} solves the Harada equation with the traceless energy-momentum tensor
\begin{align*}
T_{kl}=  \frac{1}{r_0^2}(2u_k u_l +g_{kl}) -2 \dot u_k \dot u_l +\frac{\kappa}{r_0} r^2 (u_k\dot u_l + \dot u_k u_l).  
\end{align*}
\end{ex}

\subsection{\bf Cotton gravity in De Sitter space-times}\quad\\
Constant curvature space-times are defined by the Riemann tensor
$$ R_{jklm} = \frac{R}{n(n-1)} (g_{jl}g_{km} - g_{jm} g_{kl}) $$
and include De Sitter, anti De Sitter, Milne, Lanczos space-times. 
They are conformally flat $(C_{jklm}=0)$, and Einstein ($R_{jk}=g_{jk} R/n$).\\
Harada made the remark that a De Sitter metric is a vacuum solution ($T_{jk}=0$) of the Cotton gravity equation \eqref{HEq}. 
Ferus proved that the following is the only non-trivial Codazzi tensor in a De Sitter space-time \cite{Ferus} and \cite{Besse} p.436:
\begin{align}
\mathscr C_{jk} =\nabla_j \nabla_k \phi +\frac{R\phi }{n(n-1)}g_{jk}  \label{FERUS}
\end{align} 
where $\phi $ is a smooth scalar field. Then 
\begin{align}
T_{kl} =&  \frac{R}{n}g_{kl} - \frac{R}{2} g_{kl}  -\mathscr C_{kl} + g_{kl}\mathscr C^j{}_j  \nonumber \\
&- g_{kl}\left [\nabla_j\nabla^j \phi + \frac{1}{n}R\phi -\frac{n-2}{2n}R\right ] -\nabla_k \nabla_l \phi 
\end{align}
is the most general energy-momentum tensor for Cotton gravity in a De Sitter space-time. 

We now discuss an extension. 
Consider $\mathscr C_{jk}= \nabla_j \nabla_k \phi +K \phi g_{jk} $ with $K $ constant and time-like $\nabla_k\phi$. The vector $u_k = \nabla_k\phi/\sqrt{-\nabla^j\phi \nabla_j \phi}$ is time-like unit. \\
By the Ricci identity, the condition that $\mathscr C_{jk}$ is a Codazzi tensor is 
$$R_{jkl}{}^m u_m  + K (g_{kl}u_j - g_{jk}u_k)=0.$$
Contraction with $g^{jl}$ shows that $u_m  $ is eigenvector of the Ricci tensor: $R_{k}{}^m u_m  = (n-1)K u_k $. 
In analogy with Prop.\ref{23} one evaluates the Ricci tensor:
$$ R_{jk} = \left[\frac{R}{n-1} - nK \right]u_j u_k +  \left[\frac{R}{n-1} - K \right]g_{jk} - (n-2) E_{jk} $$
where $E_{jk}$ is the electric tensor.  
With the Ricci and Codazzi tensors, the tensor
\begin{align}
T_{kl} = R_{kl} - g_{kl} \frac{R}{2}  - \nabla_j \nabla_k \phi  - g_{jk} \left [\nabla_j\nabla^j \phi + (n-1)K\phi \right ]
\end{align}
is the energy-momentum tensor for the Cotton gravity equation \eqref{HEq} with any metric such the given $\mathscr C_{jk}$ is Codazzi.

In the cases described above, the Codazzi tensor introduces a coupling of gravity with a scalar field.

\section{Conclusion}
Codazzi tensors have an intrinsic geometric importance, 
and naturally enter in the recently proposed Cotton gravity by Harada. 
The specific form of a Codazzi tensor restricts the space-time it lives in. These facts allow for a strategy to find solutions of the Cotton gravity.\\ 
We investigated two specific forms of Codazzi tensors: the perfect fluid and the current flow. In the first case the hosting metric turns out to be a generalization of Stephani Universes. In the literature, Stephani Universes are conformally flat cosmological solutions of the Einstein equations with perfect fluid source.\\ 
In the second case, a static current flow Codazzi tensor generates metrics that embrace Nariai and Bertotti-Robinson space-times,
and extensions. In the literature they are solutions of various gravity theories, such as Einstein, low energy string, 
Einstein-Maxwell and so on. By construction, all these metrics solve the Harada-Cotton gravity in geometries selected by the Codazzi tensor, with stress-energy tensors different from the original theory. \\
An interesting question is whether other forms of Codazzi tensors may give rise to new solutions of Cotton gravity 
of physical interest, using the same strategy. 

\section*{Appendix: Proof of theorem \ref{ONE}}
\begin{proof} 
For a perfect fluid tensor, the Codazzi condition $0=\nabla_i \mathscr C_{jk}-\nabla_j \mathscr C_{ik}$ is:
\begin{align}
0=&u_k (u_j \nabla_i A -u_i\nabla_j A) + (g_{jk}\nabla_i B -g_{ik}\nabla_j B) \nonumber\\
& + Au_k (\nabla_i u_j - \nabla_j u_i) + A(u_j\nabla_i u_k-u_i\nabla_j u_k) \label{Codazzi}
\end{align}
Contraction with $u^k$, and $u^k\nabla_j u_k=0$ give: $0=-u_j \nabla_i (A-B) +u_i\nabla_j (A-B)- A(\nabla_i u_j - \nabla_j u_i) $. Another contraction with $u^j$
\begin{align}
0=\nabla_i (A-B) +u_i (\dot A-\dot B) + A\dot u_i  \label{AB}
\end{align}
simplifies the previous equation to:  $0= A(u_j \dot u_i  - u_i\dot u_j- \nabla_i u_j + \nabla_j u_i) $. Since $A\neq 0$:
$0= u_j \dot u_i  - u_i\dot u_j- \nabla_i u_j + \nabla_j u_i $. By inserting the standard decomposition 
$$\nabla_i u_j = \varphi (g_{ij}+u_i u_j) +\sigma_{ij} + \omega_{ij} - u_i\dot u_j$$ 
where $\varphi $ is the expansion parameter, $\sigma_{ij} $
is the shear and $\omega_{ij}$ is the vorticity, the result is: $\omega_{ij}=0$ (the perfect fluid velocity is vorticity-free).

The contraction of \eqref{Codazzi} with $g^{jk}$ is:
\begin{align}
0= -\nabla_i A -u_i\dot A + (n-1)\nabla_i B  - A\dot u_i - Au_i \nabla_k u^k \label{ABtheta}
\end{align}
Contraction with $u^i$: $(n-1)\dot B+A \nabla_k u^k=0$. Then the equation becomes:
$$0= -\nabla_i [A- (n-1)B] -u_i[\dot A -(n-1)\dot B]  - A\dot u_i $$
Together with \eqref{AB} the equations give:
\begin{align*}
0= -\nabla_i [A- (n-1)B] -u_i[\dot A -(n-1)\dot B]  + \nabla_i (A-B)
+u_i (\dot A-\dot B)
\end{align*}
i.e. $\nabla_i B =-u_i \dot B$. Then eq.\eqref{AB} gives:
$\nabla_i A = -u_i \dot A - A \dot u_i $.

Contraction with $u^i$ of \eqref{ABtheta} gives $(n-1)\dot B + A\nabla_k u^k=0$ i.e. $\varphi = -\frac{\dot B}{A}$.\\ 
Let us introduce the expansion of $\nabla_i u_j$ with $\omega_{ij}=0$ in \eqref{Codazzi}. Several terms simplify to give: 
$u_j \sigma_{ik}  = u_i \sigma_{jk}$ i.e. $\sigma_{jk}=0$. Therefore: $\nabla_i u_j = \varphi (g_{ij} + u_i u_j ) - u_i\dot u_j$.
Finally, we evaluate the gradient of the expansion parameter:
\begin{align*}
\nabla_i \varphi &= -\frac{1}{A}\nabla_i \dot B +\dot B \frac{\nabla_i A}{A^2}\\ 
&= -\frac{1}{A} \nabla_i (u^k\nabla_k B) +\frac{\dot B}{A^2} (-u_i \dot A  - \dot u_i A)
\end{align*}
The first term contains $(\nabla_i u^k)\nabla_k B + u^k \nabla_i\nabla_k B = [\varphi (\delta_i^k + u_i u^k) - u_i \dot u^k] (-u_k \dot B) + u^k \nabla_k 
(-u_i \dot B) = -\dot u_i \dot B -u_i \ddot B$. Then:
$\nabla_i \varphi =  - u_i (\frac{\dot B \dot A}{A^2} - \frac{\ddot B}{A}) = -u_i\dot \varphi$. 

The opposite statement holds: if a perfect fluid tensor solves eqs.\eqref{C1}-\eqref{C5} then it is Codazzi. The right hand side of eq.\eqref{Codazzi} is 
evaluated with the conditions:
\begin{align*}
& u_k (u_j \nabla_i A -u_i\nabla_j A) + (g_{jk}\nabla_i B -g_{ik}\nabla_j B) \nonumber\\
& + Au_k (\nabla_i u_j - \nabla_j u_i) + A(u_j\nabla_i u_k-u_i\nabla_j u_k)\\
&=- Au_k (u_j \dot u_i  -u_i\dot u_j ) - \dot B (g_{jk}u_i  -g_{ik}u_j) \nonumber\\
&\quad  - Au_k (u_i \dot u_j - u_j \dot u_i) + A[u_j (\varphi g_{ik} -u_i\dot  u_k)\\ 
&\quad -u_i(\varphi g_{jk} - u_j \dot u_k)]\\
&=- (\dot B +\varphi A)(g_{jk}u_i  -g_{ik}u_j)=0
\end{align*}
with use of the expression \eqref{C5} for $\varphi $.
\end{proof}

\section*{Appendix: Proof of theorem \ref{Cflow}}
\begin{proof}
Suppose that the tensor $\nabla_i \mathscr C_{jk} -\nabla_j \mathscr C_{ik}=0$ is Codazzi,  with $\nabla_j\dot u_k = \nabla_k \dot u_j$.\\
1) Since $\dot u^k u_k=0$, it is $\dot u^k\nabla_j u_k = -u^k \nabla_j \dot u_k = -u^k\nabla_k \dot u_j = -\ddot u_j$.\\
2) Since $\dot u^k u_k=0$, it is $\ddot u^k  u_k = -\dot u^k \dot u_k = -\eta $.\\
3) Contraction of the Codazzi condition with $g^{jk}$ gives:
\begin{align*} 
0=&\nabla^k [\lambda (u_i \dot u_k + \dot u_i u_k)]\\
=& u_i (\dot u^p\nabla_p\lambda +\lambda \nabla_k\dot u^k ) 
+\dot u_i (\dot \lambda +\lambda\nabla_k u^k) +\lambda \ddot u_i +\lambda \dot u^k\nabla_k u_i
 \end{align*}
 Contraction with $\dot u^i$ and contraction with $u^i$, with properties (1), (2) give:
 \begin{align*}
& \dot \lambda  + \lambda \nabla_k u^k =0\\
& \dot u^p\nabla_p\lambda +\lambda \nabla_k\dot u^k  +\lambda \eta =0
 \end{align*}
What remains of the equation is $0=-\eta u_i +\ddot u_i +\dot u^k\nabla_k u_i$.\\
4) Contraction of the Codazzi condition with $u^k$ is:
\begin{align}
0&=\nabla_i \mathscr (u^k\mathscr C_{jk}) - \mathscr C_{jk} \nabla_i u^k  -\nabla_j (u^k \mathscr C_{ik})+ \mathscr C_{ik} \nabla_j u^k \nonumber\\
&= \nabla_i(-\lambda \dot u_j) -\lambda u_j \dot u_k \nabla_i u^k   \nabla_j(-\lambda \dot u_i) -\lambda u_i \dot u_k \nabla_j u^k\nonumber \\ 
&= -\nabla_i(\lambda \dot u_j) + \lambda u_j \ddot u_i + \nabla_j(\lambda \dot u_i) - \lambda u_i \ddot u_j\nonumber\\
&=-(\nabla_i \lambda) \dot u_j + \lambda u_j \ddot u_i + (\nabla_j\lambda) \dot u_i - \lambda u_i \ddot u_j \label{STAR}
\end{align}
Contraction with $u^i$ is: $0=- \dot \lambda \dot u_j + \lambda u_j u^i \ddot u_i  + \lambda  \ddot u_j$ i.e.
\begin{align*}
\lambda \ddot u_j =  \lambda \eta u_j +\dot \lambda \dot u_j 
\end{align*}
Using this in eq.\eqref{STAR} gives: 
$0=-(\nabla_i \lambda +\dot \lambda u_i) \dot u_j  + (\nabla_j\lambda + \dot \lambda u_j) \dot u_i$ with solution
\begin{align}
\nabla_i \lambda = -\dot \lambda u_i + \dot u_i \frac{\dot u^p\nabla_p \lambda}{\eta}  \label{LLL}
\end{align} 
5) Let us rewrite in full the Codazzi condition, using the results found so far.\\ 
To manage it, begin with:
\begin{align*}
\nabla_i \mathscr C_{jk}
=[-\dot \lambda u_i + \dot u_i \frac{\dot u^p\nabla_p \lambda}{\eta}]  (u_j \dot u_k + \dot u_j u_k)
+\lambda \nabla_i (u_j \dot u_k + \dot u_j u_k)
\end{align*}
Now subtract the expression with first two indices exchanged, and use closedness:
\begin{align}
0&=(\dot u_i u_j  - u_i \dot u_j)(\dot\lambda u_k +\frac{\dot u^p\nabla_p\lambda}{\eta}\dot u_k)+ \lambda (\nabla_i u_j -\nabla_j u_i)\dot u_k \nonumber\\
& +\lambda (u_j \nabla_i \dot u_k - u_i\nabla_j \dot u_k) +\lambda (\dot u_j \nabla_i u_k - \dot u_i\nabla_j u_k) 
 \label{CODA}
\end{align}
6) Contraction with $u^i$ and elimination of $\ddot u_k$:
\begin{align*}
0=&  \dot u_j (\dot\lambda u_k +\frac{\dot u^p\nabla_p\lambda}{\eta}\dot u_k)+ 2\lambda \dot u_j \dot u_k
 +\lambda (u_j  \ddot u_k + \nabla_j \dot u_k)\\ 
 =&\dot u_j (\dot\lambda u_k +\frac{\dot u^p\nabla_p\lambda}{\eta}\dot u_k)+ 2\lambda \dot u_j \dot u_k 
 +\lambda \eta u_j  u_k + \dot \lambda u_j \dot u_k + \lambda \nabla_j \dot u_k
 \end{align*}
We then obtain:
 \begin{align}
\lambda \nabla_j \dot u_k = -\lambda \eta u_j u_k -\dot\lambda (\dot u_j u_k + u_j \dot u_k) 
-\dot u_j \dot u_k (2\lambda +\frac{\dot u^p\nabla_p\lambda}{\eta}) \label{nabladot}
 \end{align}
 and the contraction with $\dot u^k$: $\frac{1}{2}\lambda \nabla_j \eta = -\dot\lambda  \eta u_j -  \dot u_j  (2\lambda\eta + \dot u^p\nabla_p\lambda )$. In particular,
 \begin{align}
2\lambda+\lambda \frac{\dot u^j \nabla_j \eta}{2\eta^2} + \frac{\dot u^p\nabla_p\lambda}{\eta} =0
 \end{align}
 This relation in \eqref{LLL} and in \eqref{nabladot} respectively gives equations \eqref{D2} and \eqref{D3}.\\ 
7) Contraction of \eqref{CODA} with $\dot u^k$ is:
\begin{align*}
0=(\dot u_i u_j  - u_i \dot u_j) (\dot u^p\nabla_p\lambda) + \lambda \eta (\nabla_i u_j -\nabla_j u_i) \\
+\tfrac{1}{2}\lambda (u_j \nabla_i \eta - u_i\nabla_j \eta ) - \lambda (\dot u_j \ddot u_i - \dot u_i\ddot u_j) 
\end{align*}
Now specify $\ddot u_k$ and $\nabla_k \eta $: $ 0=(u_i \dot u_j - u_j \dot u_i) + (\nabla_i u_j - \nabla_j u_i) $. This statement means that the velocity is vorticity-free. \\
8) Contraction of \eqref{CODA} with $\dot u^i$:
\begin{align*}
0 =&  u_j (\dot\lambda \eta u_k + \dot u_k \dot u^p\nabla_p\lambda)+ \lambda \dot u^i (\nabla_i u_j - \nabla_j u_i)\dot u_k \\
&+\lambda (u_j \dot u^i\nabla_i \dot u_k + \dot u_j \dot u^i \nabla_i u_k - \eta\nabla_j u_k)\\
=&  u_j (\dot\lambda \eta u_k + \dot u_k \dot u^p\nabla_p\lambda)+ \lambda \eta  u_j \dot u_k \\
&+\lambda (u_j \dot u^i\nabla_i \dot u_k + \dot u_j \dot u^i \nabla_i u_k - \eta\nabla_j u_k)
\end{align*}
Note that $\dot u^i\nabla_i \dot u_k = \dot u^i\nabla_k \dot u_i = \frac{1}{2}\nabla_k \eta = - (\dot\lambda /\lambda) \eta u_k -  \dot u_k  [2\eta + 
(\dot u^p\nabla_p\lambda )/\lambda ]$.\\ 
Next, a result in (3) is:  $ \dot u^i \nabla_i u_k =- \ddot u_k +\eta u_k = -(\dot \lambda/\lambda) \dot u_k$. We then obtain:
\begin{align*}
\lambda \eta \nabla_j u_k =& u_j (\dot u_k \dot u^p\nabla_p\lambda+ \lambda \eta \dot u_k) -  u_j\dot u_k  (2\eta \lambda + 
\dot u^p\nabla_p\lambda ) \\ 
& - \dot \lambda \dot u_j \dot u_k = -\lambda \eta u_j \dot u_k   -\dot \lambda \dot u_j \dot u_k
\end{align*}
Then: $\nabla_j u_k = - \frac{\dot \lambda}{\lambda} \frac{\dot u_j \dot u_k}{\eta}  - u_j \dot u_k $.

Now we prove the statement the way back. Let's evaluate with conditions \eqref{D1}-\eqref{D3} and closed $\dot u_i$:
\begin{align*}
\nabla_i \mathscr C_{jk}-\nabla_j \mathscr C_{ik} =(u_j \dot u_k + \dot u_j u_k)\nabla_i \lambda -(u_i \dot u_k + \dot u_i u_k)\nabla_j \lambda \\
+\lambda [\nabla_i(u_j \dot u_k + \dot u_j u_k) -\nabla_j (u_i \dot u_k + \dot u_i u_k)].
\end{align*}
In the first line we use \eqref{D2}:
\begin{align*}
&  -[ u_i \dot \lambda + \lambda \dot u_i (2+ \frac{\dot u^p\nabla_p\eta}{2\eta^2} )] (u_j \dot u_k + \dot u_j u_k)\\ 
&+[ u_j \dot \lambda + \lambda \dot u_j (2+ \frac{\dot u^p\nabla_p\eta}{2\eta^2} )](u_i \dot u_k + \dot u_i u_k)\\
=&-(u_i  \dot u_j - u_j \dot u_i) [u_k \dot \lambda -\lambda  \dot u_k (2+ \lambda \dot u_k \frac{\dot u^p\nabla_p\eta}{2\eta^2})].
\end{align*}
In the second line, we use closedness, \eqref{D1} and \eqref{D3}:
\begin{align*}
&\lambda [(\nabla_i u_j -\nabla_j u_i) \dot u_k + \dot u_j \nabla_i u_k -\dot u_i \nabla_j u_k +u_j \nabla_i \dot u_k - u_i\nabla_j \dot u_k]\\
&=\lambda [(-u_i \dot u_j +u_j \dot u_i) \dot u_k - \dot u_j u_i \dot u_k +\dot u_i u_j \dot u_k ]\\
&\quad +(u_i \dot u_j -\dot u_i u_j) (\dot\lambda  u_k -\lambda \dot u_k   \frac{\dot u^p\nabla_p\eta }{2\eta^2})]\\
&= (u_i \dot u_j -u_j \dot u_i) \left [ -2\lambda \dot u_k + \dot\lambda u_k -  \lambda\dot u_k \frac{\dot u^p\nabla_p\eta}{2\eta^2} \right ].
\end{align*}
The addends cancel and $\nabla_i \mathscr C_{jk}-\nabla_j \mathscr C_{ik} =0$.
\end{proof}

\centerline{Data availability}
Data sharing not applicable to this article as no datasets were generated or analysed during the current study.

\end{document}